\documentstyle[aps,pra,twocolumn]{revtex}
\begin{document}
\draft

\title{
Complete positivity of nonlinear evolution: A case study
}
\author{
Marek Czachor$^{1,2}$ and Maciej Kuna$^1$
}
\address{
$^1$Wydzia{\l}  Fizyki Technicznej i Matematyki Stosowanej\\
 Politechnika Gda\'{n}ska,
ul. Narutowicza 11/12, 80-952 Gda\'{n}sk, Poland\\
$^2$ Arnold Sommerfeld Institute for Mathematical Physics\\
Technical University of Clausthal, D-38678 Clausthal-Zellerfeld,
Germany}
\maketitle
\begin{abstract}
Simple Hartree-type equations lead to dynamics of a subsystem that is not
completely positive in the sense accepted in mathematical
literature. In the linear case this would imply that negative
probabilities have to appear for some system that contains
the subsystem in question. In the nonlinear case this does not
happen because the mathematical definition is physically
unfitting as shown on a concrete example. 
\end{abstract}

\vskip1pc

\section{Complete positivity and nonlinearity}

Linear maps that are positive but not {\it completely\/} positive (CP)
\cite{cp1,cp2,cp3,cp4,cp5,cp6,cp7,M84} have been
shown to play an essential role in characterization of degree of
entanglement of correlated quantum systems
\cite{Peres,Hor1,Hor2}, an important issue for quantum
computation and cryptography. 
A possibility of an experimental verification of CP 
of quantum evolutions was discussed 
in the context of the neutral kaon decay problem in \cite{Ben1}.
Although the notion of CP may seem somewhat abstract and technical, it
has a simple physical interpretation for {\it linear\/} maps.
We begin with
a system, labelled ``1", whose dynamics is given by
some positive map 
$\phi_1^t(a)= a(t)$, 
$\phi_1^t:\, {\cal A}\to  {\cal A}$
where ${\cal A}$ is a set of bounded operators acting  in
a Hilbert space ${\cal H}_1$. To avoid technicalities we assume
${\cal H}_1$ is finite dimensional. In linear quantum 
mechanics a reversible dynamics is given by 
$\phi_1^t(a)=U_taU_t^{-1}$ where $U_t$ is unitary. 

We require positivity of $\phi_1^t$ since if $a$
is a density matrix we want the same to be true for $a(t)$. 
Now consider a density matrix $\rho_{1+2}(0)$ of some bigger
system ``1+2" consisting of the original one plus a system whose
dimension is $m$ and which evolves trivially (its $U_t=1$). 
The initial density matrix of ``1+2" is of the form 
\begin{eqnarray}
\rho_{1+2}(0)
&=&
\sum_{ss'kl}\rho_{1+2}(0)_{ss'kl}
|s\rangle\langle s'|
\otimes
|k\rangle\langle l|\\
&=&
\sum_{k,l=1}^m a_{kl}
\otimes
|k\rangle\langle l|.\label{asss}
\end{eqnarray}
In the context of CP maps it is more standard to use the
isomorphic form \cite{iso}
\begin{eqnarray}
\rho_{1+2}(0)
&=&
\left(
\begin{array}{ccc}
a_{11} & \dots & a_{1m} \\
\vdots & \vdots & \vdots\\
a_{m1} & \dots & a_{mm}
\end{array}
\right).\label{akl}
\end{eqnarray}
It follows, the argument continues, that since the dynamics on 
${\cal A}$ is given by $\phi_1^t$ each of the entries evolves by
$a_{kl}\mapsto \phi_1^t(a_{kl})$ and the whole density matrix 
is mapped into 
\begin{eqnarray}
\rho_{1+2}(t)
&=&
\left(
\begin{array}{ccc}
\phi_1^t(a_{11}) & \dots & \phi_{1}^t(a_{1m}) \\
\vdots & \vdots & \vdots\\
\phi_{1}^t(a_{m1}) & \dots & \phi_{1}^t(a_{mm})
\end{array}
\right)
\label{big}
\end{eqnarray}
which in linear quantum mechanics reduces to 
$$
\sum_{k,l=1}^m a_{kl}
\otimes
|k\rangle\langle l|\to 
U_t\otimes 1_2
\Bigg(
\sum_{k,l=1}^m a_{kl}
\otimes
|k\rangle\langle l|
\Bigg)
U_t^{-1}\otimes 1_2.
$$
If $\rho_{1+2}(t)= \phi_{1+2}^t\big(\rho_{1+2}(0)\big)$ is to be
a density matrix it 
should not lead to negative probabilities. 
Moreover one should be able to do the
construction for any $m$. If this is the case the map $\phi_1^t$
is said to be CP. The dynamics one typically
thinks of in quantum mechanics is linear and therefore the
notion of CP was initially defined only for
linear maps \cite{T}. However there are many situations in
quantum physics where the dynamics is nonlinear. 
A nonlinear evolution of observables in Heisenberg picture is
typical of quantum optics and field theory. Nonlinearly evolving
states appear in mean field theories (Hartree-type equations
\cite{Bona}), soliton theory (nonlinear Schr\"odinger equations),
and various attempts of nonlinear generalizations of quantum mechanics.
Although the latter theories do not yet
correspond to any concrete physical situation they have led to
some formal developments especially due to the famous
``Einstein-Podolsky-Rosen malignancy" discussed by Gisin and
others \cite{Gisin1,Gisin2,Polchinski,MCfpl} (see Appendix~C).

The argument for CP we have presented does not
seem to crucially depend on the linearity of $\phi_1^t$. It is
therefore natural to extend the above definition of CP
 also to maps which are not linear. 
This was done independently by Ando and
Choi \cite{AC} and Arveson \cite{A} and a general structure
theorem characterizing all CP (linear and
nonlinear) maps was found. Apparently the problem was solved.

A surprise came when Majewski and Alicki showed in \cite{M,AM} that
a simple Hartree-type nonlinear evolution of a
finite-dimensional density matrix does not lead to a dynamics
completely positive in the sense of \cite{AC,A}. 
This can be shown as follows.
Consider a nonlinear equation
$
i\dot \rho = [h(\rho),\rho]\label{HF}
$
where $h(\rho)={\,\rm Tr\,} (Q\rho)Q/{\rm Tr\,}\rho$ 
is a time independent nonlinear
Hamiltonian operator. The solution of the equation is 
\begin{eqnarray}
\phi^t\big(\rho(0)\big)
&=&
e^{-ih(\rho)t}
\rho(0)
e^{ih(\rho)t}.\label{rhoHF}
\end{eqnarray}
To show that (\ref{rhoHF}) is not completely positive in the
sense of Ando, Choi and Arveson it is sufficient to note that 
$\phi^t(\lambda\rho)=\lambda
\phi^t(\rho)$ whereas there is a theorem \cite{AC}
stating that a completely positive and 1-homogeneous dynamics is
linear. The result seems to imply that {\it any\/} mean-field nonlinear 
evolution of a density matrix leads to negative probabilities!

Alicki and Majewski suggestion was to investigate more precisely the
problem of uniqueness of solutions leading to
non-completely-positive nonlinear evolutions. In particular they
pointed out 
that the generator given by the above nonlinear equation
is not accretive and the Cauchy problem can have different
solutions \cite{acc}. 
They did not however dare to challenge the basic definition
proposed in \cite{AC,A}.

We will now show that the Hartree-type evolution does not imply
negative probabilities because it is the basic Ando-Choi-Arveson 
definition that is physically unfitting. 
To do so we shall consider an example of the
Hartree-type evolution, essentially equivalent to the one
discussed in \cite{M,AM}. The new element we introduce is a
physically correct way of describing composite systems which
involve nonlinear 
evolutions. This subtle point was clarified in the papers by
Polchinski \cite{Polchinski} and Jordan \cite{J}, and generalized 
in \cite{MCpla,MCMK,MCprl}. 
We will first show that the form of correctly extended dynamics
differs from the one assumed in the discussion of complete
positivity although reduces to the standard expression if the
dynamics is linear. 
Next it will be shown that the definition of a 
completely positive map analyzed in \cite{AC,A} involves implicitly
an ill defined extension of nonlinear dynamics to tensor product
spaces. The physical problem turns out to be of the same type
as the one with the 
definition of dynamics of composite systems given by Weinberg in
\cite{W} . The Weinberg definition not only led to the
nowadays famous ``faster-than-light telegraph" but also 
predicted an apparently paradoxical disagreement between the
Bloch equation and Janes-Cummings approaches to two-level
systems \cite{MCpra}. A corrected description 
\cite{MCpra} showed that the paradox is a result of a wrong
formalism. It proved also that a precise way of describing
noninteracting systems leads to a meaningful dynamics when the
systems are coupled. 

\section{Example}

Consider two noninteracting systems described by Hamiltonian
functions  $H_{1}(\rho_{1})=\big({\,\rm Tr\,}_{1}h
\rho_{1})^2/ {\,\rm Tr\,}_{1}\rho_{1}$ and $H_{2}(\rho_{2})={\,\rm Tr\,}_{2}
\rho_{2}$. Here $\rho_{1}$ and $\rho_{2}$ are, respectively,
$n\times n$ and $m\times m$ density matrices. According to
general rules \cite{Polchinski,J,MCpla,MCprl} 
the Hamiltonian function of the composite system is 
\begin{eqnarray}
H_{1+2}(\rho_{1+2})
&=&
H_{1} \circ{\,\rm Tr\,}_2(\rho_{1+2})
+
H_{2}\circ{\,\rm Tr\,}_1(\rho_{1+2})\label{H12}\\
&=&
\frac{
\big({\,\rm Tr\,}_{1+2}h\otimes 1_{2}
\rho_{1+2})^2}{ {\,\rm Tr\,}_{1+2}\rho_{1+2}}
+
{\,\rm Tr\,}_{1+2}
\rho_{1+2}.
\end{eqnarray}
The main motivation for the definition (\ref{H12}) is the fact
that the Lie-Poisson dynamics of density matrices generated by
(\ref{H12}) allows for a complete separation of the two
subsystems: A reduced dynamics of a subsystem is characterized
entirely in terms of quantities intrinsic to this subsystem and
this holds for all initial conditions for $\rho_{1+2}$ and all
Hamiltonian functions $H_k$ (the most general discussion of
this problem can be found in \cite{MCprl} where an extension to
situations where no Hamiltonian function exists is also analyzed). 
The evolution is given by a Lie-Poisson 
equation  \cite{J,MCpla,B} 
$
i d\rho_{kk'}/dt=\{\rho_{kk'},H\}
$
involving, in this case, the Poisson bracket
\begin{eqnarray}
\{A,B\}
&=&
\delta_{kl'}\frac{\partial A}{\partial \rho_{kk'}} 
\rho_{lk'}\frac{\partial B}{\partial \rho_{ll'}}
- (A\leftrightarrow B)\label{PB}
\end{eqnarray}
which, when
translated into the standard matrix notation, leads to the nonlinear 
Liouville-von Neumann equations $\dot \rho_{2}=0$ and
\begin{eqnarray}
i\dot \rho_{1}
&=&
2\frac{{\,\rm Tr\,}_{1}h\rho_{1}}{{\,\rm
Tr\,}_{1}\rho_{1}}[h,\rho_{1}] \label{10}\\
i\dot \rho_{1+2}
&=&
2\frac{{\,\rm Tr\,}_{1+2}h\otimes 1_2\rho_{1+2}}{{\,\rm Tr\,}_{1+2}\rho_{1+2}}
[h\otimes 1_2,\rho_{1+2}].\label{12}
\end{eqnarray}
Equation (\ref{12}) is a very natural extension of the 1-particle
dynamics (\ref{10}) and could be taken for granted even without the general
background we have given above.  
Define 
\begin{eqnarray}
U_t\big(\rho_1(0)\big)
=
\exp\big[-2i{\,\rm Tr\,}_{1}\big(h\rho_{1}(0)\big)ht/{\,\rm
Tr\,}_{1} \rho_{1}(0)\big] 
\end{eqnarray}
All the expressions
involving traces are time independent (as depending on
Hamiltonian functions and  ${\,\rm Tr\,}
\rho$ which is a Casimir invariant). Therefore we can
immediately write the solutions $\rho_{2}(t)=\rho_{2}(0)$ and
\begin{eqnarray}
\rho_{1}(t)
&=&
U_t\big(\rho_1(0)\big)
\rho_{1}(0)
U^{-1}_t\big(\rho_1(0)\big)
\label{5}\\
\rho_{1+2}(t)
&=&
U_t\big(\rho_1(0)\big)
\otimes 1_2
\rho_{1+2}(0)
U_t\big(\rho_1(0)\big)
\otimes 1_2.
\end{eqnarray}
It is clear that the self-consistency condition
\begin{eqnarray}
{\,\rm Tr\,}_2 \circ \phi^t_{1+2}= \phi^t_{1}\circ {\,\rm Tr\,}_2\label{OK}
\end{eqnarray}
typical of a well defined dynamics is fulfilled.
It should be stressed that (\ref{OK}) is not accidental but 
follows from the very 
construction of the Lie-Poisson dynamics \cite{MCprl}.

The dynamics given by $\phi_1^t$ is nonlinear but 1-homogeneous.
The Theorem~4 in \cite{AC} states that the dynamics can not be
completely positive. It is obvious, however, that our 
dynamics preserves positivity of $\rho(t)$ both for the
subsystem and the composite system (this is a general property
of this formalism, see \cite{MCMM}). The dynamics can be uniquely
extended from subsystems to the composite ones and then again
reduced to subsystems giving the correct result, and this is of
course valid for any $m$. So the dynamics
{\it looks\/} completely positive! 

To understand what goes wrong
consider a more detailed example. Let us take the positive
matrix  as a $t=0$ density matrix (cf. the proof of Theorem~4 in \cite{AC}):
\begin{eqnarray}
\rho_{1+2}(0) 
&=&
\left(
\begin{array}{cccc}
a & a & a & a \\
a & a+b & a+b & a\\
a & a+b & a+b & a\\
a & a & a & a
\end{array}
\right)
\end{eqnarray}
where $a$, $b$ are positive and Hermitian $n\times n$ matrices
(so here we take $m=4$). A reduced density matrix corresponding
to the nonlinear subsystem is 
$
\rho_{1}(0) ={\,\rm Tr\,}_2 \rho_{1+2}(0) =
4a + 2b.
$
The solution for the subsystem is 
\begin{eqnarray}
\rho_{1}(t)
&=&
U_t(2a+b)
(4a + 2b)
U^{-1}_t(2a+b)\label{16}
\end{eqnarray}
The solution for the whole system is
\begin{eqnarray}
{}&{}&\rho_{1+2}(t)\nonumber\\
&{}&=
\tilde U_t(2a+b) 
\left(
\begin{array}{cccc}
a & a & a & a \\
a & a+b & a+b & a\\
a & a+b & a+b & a\\
a & a & a & a
\end{array}
\right)
\tilde U^{-1}_t(2a+b)
\label{right},
\end{eqnarray}
where $\tilde U_t(2a+b)=U_t(2a+b) \otimes 1_2$.
This dynamics is consistent 
with (\ref{16})
but is not in the form one assumes
in \cite{AC,A,M,AM}! Indeed what one
typically assumes would correspond to
\begin{eqnarray}
\rho_{1+2}(t)=
\left(
\begin{array}{cccc}
\phi_1^t(a) & \phi_1^t(a) & \phi_1^t(a) & \phi_1^t(a) \\
\phi_1^t(a) & \phi_1^t(a+b) & \phi_1^t(a+b) & \phi_1^t(a)\\
\phi_1^t(a) & \phi_1^t(a+b) & \phi_1^t(a+b) & \phi_1^t(a)\\
\phi_1^t(a) & \phi_1^t(a) & \phi_1^t(a) & \phi_1^t(a)
\end{array}
\right).\label{wrong}
\end{eqnarray}
It is sufficient to compare the ``11" entries of
(\ref{wrong}) and (\ref{right}) to see that they are different. 
The correct dynamics (\ref{right}) gives 
\begin{eqnarray}
a \to 
U_t(2a+b)
a
U^{-1}_t(2a+b)
\end{eqnarray}
whereas (\ref{wrong}), which one naively expects, would give 
\begin{eqnarray}
a \to 
U_t(a)
a
U^{-1}_t(a)
\end{eqnarray}
Actually, it can be shown that a physically correct dynamics {\it
cannot\/} be in the form (\ref{wrong}) for two essential
reasons. 
Indeed, let us first
note that the bases chosen in (\ref{asss}) are arbitrary. 
Choosing a new basis $|\tilde k\rangle$ in ``2" instead of $|k\rangle$
one obtains 
\begin{eqnarray}
\rho_{1+2}(0)
&=&
\sum_{ss'\tilde k\tilde l}
\tilde \rho_{1+2}(0)_{ss'\tilde k\tilde l} 
|s\rangle\langle s'|
\otimes
|\tilde k\rangle\langle \tilde l|\\
&=&
\sum_{\tilde k,\tilde l=1}^m \tilde a_{\tilde k\tilde l}
\otimes
|\tilde k\rangle\langle \tilde l|.\label{1asss}
\end{eqnarray}
The new ${\cal A}$-valued matrix 
\begin{eqnarray}
\left(
\begin{array}{ccc}
\tilde a_{11} & \dots & \tilde a_{1m} \\
\vdots & \vdots & \vdots\\
\tilde a_{m1} & \dots & \tilde a_{mm}
\end{array}
\right)\label{1akl}
\end{eqnarray}
is related to (\ref{akl}) by a similarity transformation 
$
\tilde a_{kl}=\sum_{k'l'}U_{kk'} a_{k'l'}U^{-1}_{l'l}
$
where $U_{kk'}$ is a $\bbox C$-valued unitary  $m\times m$ matrix. 

In the generic case the choice of bases is arbitrary and no
physically meaningful quantity in the subsystem ``1" can depend
on the choices made in ``2". Mathematically this means that all
physical quantities in ``1" should be invariant under
unitary similarity transformations of (\ref{akl}) or, which is
equivalent, changes from (\ref{asss}) to (\ref{1asss}). If the
latter condition is not satisfied the density matrix 
$\phi^t_{1+2}\big(\rho_{1+2}(0)\big)$ is mathematically ill defined for
$t>0$. A knowledge of $\rho_{1+2}(0)$ is insufficient for
predicting $\rho_{1+2}(t)$: One has to additionally fix a basis.
It is obvious that this problem does not occur for the
``correctly" extended dynamics. 

(\ref{wrong}) implies that after time $t$ the reduced density 
matrix is
\begin{eqnarray}
\rho_1(t) 
&=&
\phi_1^t(a)
+\phi_1^t(a+b)
+\phi_1^t(a+b)
+\phi_1^t(a).\label{wr1}
\end{eqnarray}
Assume that at $t=0$ we change
the basis in ``2" by the unitary transformation
\begin{eqnarray}
\left(
\begin{array}{cccc}
1 & 0 & 0 & 0\\
0 & 1 & 0 & 0\\
0 & 0 & \frac{1}{\sqrt{2}} & -\frac{1}{\sqrt{2}}\\
0 & 0 & \frac{1}{\sqrt{2}} & \frac{1}{\sqrt{2}}
\end{array}
\right).
\end{eqnarray}
Let us stress again the important fact that this unitary
transformation represents a {\it passive\/} modification of
coordinates of $\rho_{1+2}(0)$ resulting from the change of
basis. This should not be confused with the {\it active\/} 
transformation $\rho_{1+2}(0)\to \bbox 1\otimes U \rho_{1+2}(0)
\bbox 1\otimes U^{-1}$. Such a transformation would in general change
$\rho_{1+2}(0)$ into a new density matrix $\rho'_{1+2}(0)$ and
it might not be very surprising that two different density
matrices evolve differently. The transformations we discuss
leave $\rho_{1+2}(0)$ unchanged and still change the dynamics.
Indeed, the reduced dynamics of ``1" becomes 
\begin{eqnarray}
\tilde \rho_1(t) 
&=&
\phi_1^t(a)
+\phi_1^t(a+b)
+\phi_1^t(\frac{b}{2})
+\phi_1^t(2a + \frac{b}{2}).\label{wr2}
\end{eqnarray}
Obviously the reduced dynamics of ``1" is now different and this
is essentially the celebrated ``faster-than-light telegraph" of
Gisin \cite{Gisin1,MCcesena}. In the original Gisin telegraph
described in \cite{Gisin1} one performs the change of basis in
``2" by changing the direction of a Stern-Gerlach device.
Assuming that each measurement of spin in ``2" reduces the
two-particle entangled state to a concrete eigenstate in ``1"
one nonlocally decomposes the beam of particles into two
sub-beams which are assumed to evolve independently.
Mathematically this amounts to assuming that the reduced density
matrix in ``1" is a convex combination of projectors
corresponding to the chosen basis. 
The effect is based on the fact that for a nonlinear map
$\phi_1^t$ and two different ways of writing the density matrix
of ``1" as convex combinations
$
\rho_1(0)=
\sum_k p_k \varrho_k=\sum_k \tilde p_k \tilde \varrho_k
$
one has 
$
\sum_k p_k \phi_1^t(\varrho_k)\neq\sum_k \tilde p_k
\phi_1^t(\tilde \varrho_k).
$
The reduced density matrices of ``1" we obtain by the reductions
(\ref{wr1}) and (\ref{wr2}) have these properties. 
To explicitly see that they are different 
take $h=\sigma_z$, $a=\frac{1}{16}\big(\bbox 1
+\sigma_x\big)$, $b=\frac{1}{8}\big(\bbox 1
+\sigma_z\big)$, where $\sigma_k$ are the Pauli matrices. Then
\begin{eqnarray}
\rho_1(t) -\tilde\rho_1(t) 
&=&
-
\frac{1}{4}
\sin^2\frac{2}{3}t 
\Bigg[
\cos \frac{4}{3}t
\sigma_x +
\sin \frac{4}{3}t
\sigma_y
\Bigg].
\end{eqnarray}
The correct dynamics is free of this problem because the
nonlinear terms occuring in the reduced
density matrix are basis independent. 

Paraphrasing Gisin's statement \cite{Gisin3} one can say that a nonlinear
evolution which is completely positive in the sense of
\cite{AC,A} is physically relevant if and only if it is {\it
linear\/}. A few years ago this might be a perfect argument
against nonlinear quantum mechanics. 

The lesson we are taught by the example is the
following. First, to speak about the composition problem in
nonlinear theories, one has to specify the way the subsystems
``1" and ``2" 
evolve. This concerns not only the subsystem ``1" we are interested
in, but also the ``rest" (this, in principle, can also be a
nonlinear evolution). Then one has to specify the dynamics
of the composite ``1+2" system. This is the most delicate point
and one cannot just take  {\it any\/} linear definition and use
it for a nonlinear system.  One must make sure the
definitions are basis independent if the choice of bases is
physically irrelevant (i.e., if there is no superselection rule).
For a nonlinear positive dynamics the conditions 
$
{\,\rm Tr\,}_2 \circ \phi^t_{1+2}=\phi^t_{1}\circ {\,\rm Tr\,}_2, 
$
$
{\,\rm Tr\,}_1 \circ \phi^t_{1+2}= \phi^t_{2}\circ {\,\rm Tr\,}_1,
$
play a physical role analogous to the requirement of CP for
linear maps. 
The definition of CP
accepted in \cite{AC,A} does not satisfy these requirements and
therefore the fundamental problem of a general 
characterization of physically relevant
nonlinear completely positive maps is still open. 

Our work is a part of the joint Polish-Flemish project 007. We
are grateful to W.~A.~Majewski and P.~Horodecki for comments. 
M.~C. wants to thank prof.~H.-D.~Doebner for his hospitality at the
Arnold Sommerfeld Institute, where this work was completed, and
Deutscher Akademischer Austauschdienst (DAAD) for support. 

\section{Appendix: Controversial issues}

In our analysis we have used several techniques and made some
statements that may appear controversial. 
The Appendix addresses three mutually related groups of such problems:
Definition of mixed states, projection postulate and
faster-than-light effects.  

\subsection{Mixed states}

In linear quantum mechanics mixed states are defined in several
equivalent ways. In nonlinear quantum mechanics the definitions are no
longer equivalent. 

According to the first definition a mixed state is a probability
measure on the set of pure states. This definition is 
nonunique when one switches to nonlinear observables.
To see this consider
a state vector $|\psi\rangle$ representing a spin-1/2 particle.
Its Hilbert space is ${\bbox C}^2$ and the basis vectors are
denoted by $|0\rangle$ and $|1\rangle$. A state of mixed
polarization can be represented by a ${\bbox C}^2$-valued random
variable 
\begin{eqnarray}
|\psi_{\theta}\rangle=\psi_0|0\rangle+
e^{i\theta}\psi_1|1\rangle,
\end{eqnarray}
where $\theta\in [0,2\pi)$ is a random phase. In linear quantum
mechanics the average of an observable $\hat A$ could be calculated
as follows 
\begin{eqnarray}
\frac{1}{2\pi}\int_0^{2\pi}d\theta\,
\langle\psi_{\theta}|\hat A|\psi_{\theta}\rangle
=
|\psi_0|^2
\langle0|\hat A|0\rangle
+
|\psi_1|^2
\langle1|\hat A|1\rangle,
\end{eqnarray}
which is equivalent to representing the state by
the projector-valued random variable: With probability
$|\psi_0|^2$ one finds $|0\rangle\langle 0|$, 
and   $|1\rangle\langle 1|$ with probability
$|\psi_1|^2$. When it comes to nonlinear quantum mechanics the two
approaches are inequivalent. Consider a generalized average
(i.e. observable)
\begin{eqnarray}
A(|\psi\rangle,\langle\psi|)=
(\psi_1+\psi_1^*)^5.
\end{eqnarray}
A mixture of pure states can be represented by the
``$|\psi\rangle$-valued" random variable $\theta\mapsto
|\psi_{\theta}\rangle$ and the average by 
\begin{eqnarray}
\frac{1}{2\pi}\int_0^{2\pi}d\theta\,
A(|\psi_{\theta}\rangle,\langle\psi_{\theta}|).
\end{eqnarray}
However, the mixture cannot be represented by a 
``$|\psi\rangle\langle\psi|$-valued" random variable because 
$A(|\psi\rangle,\langle\psi|)$ cannot be written as a function
of $|\psi\rangle\langle\psi|$. [{\it Proof\/}: Assume there
exists a function
$B(|\psi\rangle\langle\psi|)=
A(|\psi\rangle,\langle\psi|)$ for any $\psi$. 
But $B(|\psi\rangle\langle\psi|)=
B(|e^{i\alpha}\psi\rangle\langle e^{i\alpha}\psi|)$ for any
$\alpha$, and $A(|\psi\rangle,\langle\psi|)\neq
A(|e^{i\alpha}\psi\rangle,\langle e^{i\alpha}\psi|)$
for almost all $\alpha$. Contradiction.] 
The assumption that all nonlinear observables on pure states 
can be written as functions (or functionals) of
$|\psi\rangle\langle\psi|$ is therefore a serious restriction
(see below).
The approach to mixtures via probability measures on pure states
was developed the works of Mielnik \cite{Mielnik} who
was also the first to seriously address the question of mixtures vs.
nonlinearity (see also \cite{Bugajski}).

A definition of a mixed state which is widely used in the
literature is the following: A
mixed state is an operator $\rho$ which is Hermitian, positive, 
trace-class and normalized (${\rm Tr\,}\rho=1$), and which is
not a projector ($\rho^2\neq \rho$). This is the definition we
use in the paper. In the nonlinear framework we use a density
matrix $\rho$ is a fully quantum object and has an onthological
status analogous to this of a wave function in standard quantum
mechanics. For a modern discussion of density matrices from such
a perspective see \cite{AA}

From what we have written it does not yet follow how to
introduce dynamics. Starting with pure states $|\psi\rangle$ one
can take a nonlinear Schr\"odinger dynamics $t\mapsto
|\psi(t)\rangle$. Starting with pure states $|\psi\rangle\langle\psi|$ one
can take a nonlinear Liouville-von Neumann dynamics $t\mapsto
|\psi(t)\rangle\langle\psi(t)|$. These approaches are, in
general, inequivalent because not all Hamiltonian functions 
$H(|\psi\rangle,\langle\psi|)$ can be written as 
$H(|\psi\rangle\langle\psi|)$. However, once we have an
$H=H(|\psi\rangle\langle\psi|)$ we can treat it as a restriction
to projectors of a more general $H=H(\rho)$ and define a
Lie-Poisson dynamics in terms of the B\'ona-Jordan Poisson
bracket. This is what we do in the paper. 

Suppose now we have a dynamics of $\rho$ which preserves
Hermiticity, trace-class property, and positivity of $\rho(t)$. 
The eigenvalues of $\rho(t)$ play then a role of probabilities
analogous to those we discussed above. In spite of this expected
property of $\rho(t)$ our density matrix cannot satisfy an
ordinary convexity principle: A convex combination of two
solutions is no longer a solution of the nonlinear evolution
equation. Typically this is regarded as an argument against
nonlinearly evolving $\rho$. This problem was discussed by B\'ona
and Jordan. They proposed the following interpretation. There
are two kinds of density matrices in quantum mechanics. One
class corresponds to a situation where an experimentalist
controls the mixture by, for example, introducing the random
phase. Then the pure-state
componets of the mixture should be treated separately and the
dynamics is nonlinear at the pure state level. This also assumes
that there exists a privileged set of observables which is
controlled during an experiment. As a result what one gets is a
kind of a superselection principle. The second class of density
matrices consists of those that arise because of some reduction
procedure and entanglement. The typical example is a
one-particle subsystem of an EPR pair. An observer at one side
of this experiment has no way to control the mixture at the other
side (see below). One can add that there exists a third class of mixtures
that cannot be controlled either: These are simply very large
systems. It is impossible to control the pure-state components
of nonlinearly evolving mixtures that occur in Bose-Einstein
condenstion of atomic clouds or chemical reactions described by
nonlinear thermodynamics.  

Finally, let us give the fourth example. The nonlinear gauge
transformations introduced by Doebner and Goldin \cite{DG} are
based on the assumption that all actual measurements are based
on those of the position observable. All theories that lead to the same
probability densities in position space at any time and for all
physical situations are therefore regarded as physically
equivalent. Nonlinear gauge transformations do not change the
position space probability density and although transform a linear
Schr\"odinger dynamics into a nonlinear one, they nevertheless do
not generate any new physics \cite{method}. An extension of Doebner-Goldin
transformations to density matrices \cite{MCprarc} leads to the requirement
that the diagonal elements of density matrices in position
space, $\rho(x,x)$, must be unchanged by gauge transformations. 
Repeating the Doebner-Goldin argument one obtains a class of
nonlinear theories that are fully equivalent to the linear
Liouville-von Neumann dynamics. Such theories do not satisfy the
ordinary convexity principle but only a ``convexity principle on
the diagonal" in position space.

\subsection{Projection postulate}

Consider a solution $|\psi\rangle$ of a linear Schr\"odinger equation. 
Projection of $|\psi\rangle$ on an eigenstate of a self-adjoint
operator $\hat A$ does not pose any problem
since the projected state is again a solution of the same
equation. When one tries to perform the same operation with a nonlinear
Schr\"odinger equation one immediately faces two difficulties. 
First, the projected state is no longer a
solution of the same equation and one has to add something not
to leave the Hilbert space. This property was even used to test
the logarithmic nonlinearity \cite{bbm,Shimony,Zeilinger}. 
This ``something" one has to add may be highly nontrivial
(for example, a nonlinear gauge transformation \cite{Lucke}).
Second, if the observable one
measures is nonlinear the notion of an eigenstate is ambiguous
\cite{MCpra,miel97}. To make an argument based on the projection
postulate physically sound one has to explicitly address these issues. 
Otherwise the argument is hand-waving and cannot be conclusive.
To avoid such dilemmas it is best to use any interpretation of quantum
mechanics which is not based on the postulate. 
This is clearly acceptable and is not a pecularity of nonlinear
quantum mechanics. Similar problems occur in quantum
cosmology.

\subsection{Faster-than-light signals}

The problem with faster than light signals was independently discovered
Gisin \cite{Gisin1}, Polchinski (see the
footnote in \cite{W}),
Svetlichny \cite{Svet} and one of us \cite{MC88}. The first
paper where a possibility of a conflict between locality and
linearity was mentioned is the work of Haag and Bannier \cite{HB}.

The original formulation due to Gisin made an explicit use of
the projection postulate and was apparently model independent. An
alternative ``model-independent" version was given in
\cite{MC88}. 
The argument of Polchinski referred to the concrete
version of nonlinear quantum mechanics proposed by Weinberg. 
The Weinberg model can serve as an illustration of all the three
ways of generating the effect. It simultaneously shows what can be
done to avoid it. 

Consider two separated systems described  by  Hamiltonian   functions 
$H_1(|\varphi\rangle\langle\varphi|)=E_1\langle \varphi|\varphi\rangle$
and 
$$
H_2(|\chi\rangle\langle\chi|)=E_2\langle \chi|\chi\rangle 
+\epsilon{\langle \chi|\sigma_z|\chi\rangle ^2\over\langle
\chi|\chi\rangle }.
$$
The element which is responsible for the faster-than-light
effects is the particular form of Hamiltonian function of the
composite system which is chosen as
\begin{eqnarray}
H_{1+2}(|\psi\rangle,\langle \psi|)=
\sum_lH_1(|\varphi_l\rangle\langle\varphi_l|)+
\sum_kH_2(|\chi_k\rangle\langle\chi_k|),\label{Wcomposite}
\end{eqnarray}
where $|\varphi_l\rangle=\sum_k\psi_{kl}|k\rangle$,
$|\chi_k\rangle=\sum_l\psi_{kl}|l\rangle$. Notice that
$H_{1+2}(|\psi\rangle,\langle\psi|)\neq H_{1+2}(|\psi\rangle\langle
\psi|)$. 
There exist entangled solutions of the corresponding nonlinear 2-particle
equation 
\begin{eqnarray}
\psi=
|\varphi_1\rangle\otimes|\chi_1\rangle
+|\varphi_2\rangle\otimes|\chi_2\rangle
\label{gensol}
\end{eqnarray}
where $|\varphi_j\rangle$ and $|\chi_i\rangle$ are some solutions of the
1-particle Schr\"odinger equations corresponding to the
subsystems. 
The nonlinearity in ``2" can make
$\langle\chi_1|\chi_2\rangle$ time dependent and proportional to
$\sin(4\epsilon\langle\sigma_z\rangle t)$, where
$\langle\sigma_z\rangle$ is, in general, nonvanishing. 
Now consider the {\it linear\/} subsystem. Its reduced density
matrix obtained from (\ref{gensol}) 
contains $\langle\chi_1|\chi_2\rangle$ which depends on
$\epsilon$. As a result there exists an observable in the linear
system whose average value depends on $\epsilon$. For example
$
\langle \sigma_y\rangle ={\rm Im}\langle\chi_1|\chi_2\rangle.
$
The telegraph so obtained allows one to send information from the
nonlinear system to the linear one. A physical interpretation of
this phenomenon was given in detail in \cite{MCfpl}. 
Technically the effect follows from the fact that an appropriate 2-particle
Poisson bracket does not vanish: 
$\{\langle\sigma_y\rangle,H_2\}\neq 0$. The two functions
appearing in this bracket correspond to different subsystems. An
observation that in Weinberg's nonlinear quantum mechanics such
brackets may not vanish is due to Polchinski. The proof
given in \cite{MC88} was based on the observation that entangled
solutions of 2-particle nonliner Schr\"odinger equations may
involve states whose scalar product
$\langle\chi_1|\chi_2\rangle$ is not conserved by the dynamics. 
As such it did not, apparently, refer to a concrete model. 
It turned out, however, that although the nonconservation of
scalar products is a general property of nonlinear evolutions in
Hilbert spaces \cite{mob} the existence of appropriate entangled solutions
is not at all general. 

The telegraph described by
Gisin in \cite{Gisin1,Gisin2} works in the opposite direction
and its mathematical origin is different. The element which is
technically responsible for the Gisin effect is the basis
dependence of the Hamiltonian function (\ref{Wcomposite}). 
To see this consider two bases in the {\it linear\/} system: A
basis $|\pm\rangle$ (spin ``up" or ``down"), and some other
basis $|\alpha,\beta,\pm\rangle=U(\alpha,\beta)|\pm\rangle$ 
obtained from the ``up-down" one by means of an $SU(2)$
transformation 
$
U(\alpha,\beta)=\left(\begin{array}{cc}
\alpha &\beta\\
-\bar \beta&\bar \alpha
\end{array}
\right).
$
The 2-particle solution which is the singlet at $t=0$, written in the basis
$|\alpha,\beta,r\rangle|s\rangle$, $r,s=\pm$, is
\begin{eqnarray}
|\psi\rangle =\frac{1}{\sqrt{2}}
e^{-i (E_1+E_2-\epsilon X^2)t}
\left(\begin{array}{cc}
-\beta e^{-i  2\epsilon Xt} & \alpha e^{i  2\epsilon Xt}\\
-\bar \alpha e^{i  2\epsilon Xt} & - \bar \beta e^{-i  2\epsilon Xt}
\end{array}\right)\nonumber
\end{eqnarray}
where $X=|\beta|^2-|\alpha|^2$. 
The reduced density matrix of the {\it nonlinear\/} system
``2" is
\begin{eqnarray}
\rho_2&=&\frac{1}{2} {\bf 1}+{\rm Re} (\bar \alpha\beta) 
\sin \bigl(4\epsilon(|\alpha|^2-|\beta|^2)t\bigr)\sigma_y\nonumber\\
&\phantom =&
\phantom{\frac{1}{2} {\bf 1}}
+
{\rm Im} (\bar \alpha\beta) 
\sin \bigl(4\epsilon(|\alpha|^2-|\beta|^2)t\bigr)\sigma_x.
\end{eqnarray}
The average of $\sigma_y$ in the  nonlinear system is 
\begin{eqnarray}
\langle\sigma_y\rangle=2{\rm Re} (\bar \alpha\beta) 
\sin \bigl(4\epsilon(|\alpha|^2-|\beta|^2)t\bigr)
\end{eqnarray}
and, hence, depends on the
choice of basis made in the {\it linear\/} one. 

As we can see we have obtained the telegraphs without any use of
the projection postulate. 
Interpreting $H_{1+2}$ as an
average energy, and taking into account that its value is basis
dependent, we can conclude that this kind of description cannot
correspond to a closed system. 

It is obvious how to eliminate both phenomena. First, one has to
guarantee that the Hamiltonian function is basis independent
(this eliminates the Gisin effect). The other effect is
eliminated if any two functions corresponding to the two
subsystems commute with respect to the 2-particle Poisson
bracket. 

To make sure that the first condition is satisfied one can
require that each subsystem observable is a function of the
reduced density matrix of this subsystem. This leads naturally
to a density matrix formalism but can be done also for pure
states by restricting all 2-particle density matrices to
projectors. It is quite remarkable that the restriction of
all 1-particle observables to functions of local density
matrices  turns out to automatically
guarantee the commutability of separated observables. 
Denote by ${\rm Tr}_1$ and ${\rm Tr}_2$ the partial traces.
Consider two functions 
\begin{eqnarray}
A&=&A(\rho_{1+2})=A_1\circ {\rm
Tr}_2(\rho_{1+2})=A_1(\rho_1),\nonumber\\ 
B&=&B(\rho_{1+2})=B_2\circ {\rm
Tr}_1(\rho_{1+2})=B_2(\rho_2),\nonumber
\end{eqnarray}
and let $\{\cdot,\cdot\}$ be a 2-particle bracket. Polchinski
and Jordan noticed that then $\{A,B\}=0$. Introducing the
Casimir invariant $C_2={\rm Tr\,}(\rho^2)$ one can show that the
Poisson bracket $\{\cdot ,\cdot\}$ is a particular case
of a Nambu-type 3-bracket \cite{Mor,MCpla}: 
$\{\cdot ,\cdot\}=\{\cdot ,\cdot,\frac{1}{2}C_2\}$. 
The most general version of the Polchinski-Jordan theorem was
proved in \cite{MCpla} in the following form:

{\it Theorem\/}: Assume $A$ and $B$ are observables which are
functionals of reduced density matrices of two separated $N$-
and $M$-particle systems. Then $\{A,B,\,\cdot\,\}=0$, where the
bracket corresponds to the composite $(N+M)$-particle system.

This result allows for an extension of the above construction
to a more general class of ``Lie-Nambu" theories where instead
of $C_2$ one puts a general nonlinear function $F$ (the
commutability of observables is independent of the choice of
$F$). 

In this context we would like to add two comments. First, the
formalism of Polchinski, B\'ona and Jordan leads ultimately to
integro-differential equations, which contradicts the belief
that local physics must involve ``local" equations (for a
discussion see \cite{MCprl}). Second, there exist global
phenomena which do not necessarily lead to faster-than-light
telegraphs but have no counterpart in liner quantum mechanics
(e.g. the threshold effects discussed in \cite{GS} and the ``Big
Brother effect" occuring for a class of Lie-Nambu equations \cite{MCPey}.)


\begin{references}
\bibitem{iso}Note for math purists: The isomorphism is
$I_{\cal B}: {\cal A}\otimes M_m(\bbox C)\to M_m({\cal A})$. 
$I_{\cal B}$ depends on the choice of the basis ${\cal B}$. 
Take two different bases. The problem with the 
Ando-Choi-Arveson map $\phi:M_m({\cal A})\to M_m({\cal A})$ is
that the diagram 
$$
\begin{array}{ccccccc}
{\cal A}\otimes M_m(\bbox C) & \stackrel{I_{\cal B}}{\rightarrow} &
M_m({\cal A}) & \stackrel{\phi}{\rightarrow} &
M_m({\cal A}) & \stackrel{I_{\cal B}^{-1}}{\searrow}
& \\
 & \stackrel{I_{{\cal B}'}}{\searrow} &
M_m({\cal A}) & \stackrel{\phi}{\rightarrow} &
M_m({\cal A}) & \stackrel{I_{{\cal B}'}^{-1}}{\rightarrow}
& {\cal A}\otimes M_m(\bbox C)
\end{array}
$$
is noncommutative.
\bibitem{acc}This statement seems to apply to the
nonuniqueness of $U\big(t,\rho(0)\big)=\exp\big[-i
h\big(\rho(0)\big)t\big]$, which is always determined up to a
transformation $V_{\rho(0)}$ stabilizing $\rho(0)$: $V_{\rho(0)}\rho(0)
V_{\rho(0)}^{-1}=\rho(0)$, see \cite{Bona,B}.
\bibitem{method}We propose the following methodological
principle (``Doebner-Goldin razor"): If you discover an
impossibility principle stating that any nonlinear modification
of quantum mechanics is impossible, first perform the following test:
Take a nonlinearly gauge transformed Schr\"odinger or Liouville-von Neumann
equation. If you use a Doebner-Goldin transformation with the
parameter $\Lambda=1$ then the transformed equation will have
the ordinary kinetic and potential terms but in addition some
nonlinearity will appear. Now use this nonlinear dynamics as an
example you tried to rule out by your theorem.
\end{references}
\end{document}